\documentclass[a4paper]{article}
\usepackage[all]{xy}\usepackage[latin1]{inputenc}        %accents
\usepackage[dvips]{graphics,graphicx}
\usepackage{amsfonts,amssymb,amsmath,color,mathrsfs, amstext}
\usepackage{amsbsy, amsopn, amscd, amsxtra, amsthm,authblk}
\usepackage{enumerate,algorithmicx,algorithm}
\usepackage{algpseudocode}
\usepackage{upref}
\usepackage{geometry}
\usepackage[displaymath]{lineno}
\usepackage{float}
\usepackage{yhmath}
\usepackage{booktabs}
\usepackage{bm}
\usepackage{multirow}
\usepackage{makecell}
\usepackage[colorlinks,
            linkcolor=red,
            anchorcolor=red,
            citecolor=red
            ]{hyperref}

\numberwithin{equation}{section}

\def\b{\boldsymbol}

\def\Ai{\mathrm{\AA}}

\DeclareMathOperator{\erf}{erf}
\DeclareMathOperator{\erfc}{erfc}

\newcommand{\E}{\mathbb{E}}
\newcommand{\R}{\mathbb{R}}

\newcommand{\bbZ}{\mathbb{Z}}

\newcommand{\cO}{\mathcal{O}}

\newtheorem{theorem}{Theorem}

\newtheorem{proposition}{Proposition}

\begin{document}

\title{A random batch Ewald method for particle systems with Coulomb interactions}

\author[1,2]{Shi Jin\thanks{shijin-m@sjtu.edu.cn}}
\author[1,2]{Lei Li\thanks{leili2010@sjtu.edu.cn}}
\author[1,2]{Zhenli Xu\thanks{xuzl@sjtu.edu.cn}}
\author[1]{Yue Zhao\thanks{sjtu-15-zy@sjtu.edu.cn}}
\affil[1]{School of Mathematical Sciences, Shanghai Jiao Tong University, Shanghai, 200240, P. R. China}
\affil[2]{Institute of Natural Sciences and MOE-LSC, Shanghai Jiao Tong University, Shanghai, 200240, P. R. China}

\date{}
\maketitle

\begin{abstract}
We develop a random batch Ewald (RBE) method for molecular dynamics simulations of particle systems with long-range Coulomb interactions,
which achieves an $\cO(N)$ complexity in each step of simulating the $N$-body systems.
The RBE method is based on the Ewald splitting for the Coulomb kernel with a random ``mini-batch'' type technique introduced to
speed up the summation of the Fourier series for the long-range part of the splitting. Importance sampling is employed
to reduce the induced force variance by taking advantage of the fast decay property of the Fourier coefficients.
The stochastic approximation is unbiased with controlled variance. Analysis for bounded force fields
gives some theoretic support of the method. Simulations of two typical problems of charged systems are presented to
illustrate the accuracy and efficiency of the RBE method  in comparison to
the results from the Debye-H\"uckel theory and the classical Ewald summation,
demonstrating that the proposed method has the attractiveness of being easy to implement with the linear
scaling and is promising for many practical applications.


\end{abstract}

\section{Introduction}
Molecular dynamics simulation is among the most popular numerical methods at the molecular or atomic level to understand dynamical and equilibrium properties of many-body particle systems in many areas such as chemical physics, soft materials and biophysics \cite{ciccotti1987simulation,frenkel2001understanding,FPP+:RMP:2010}.
However, the long-range interactions such as electrostatic Coulomb interactions pose a major challenge to particle simulations,
as one has to take into account all pairs of interactions, leading to $\cO(N^2)$ computational cost per iteration for naive discretizations,
which is not only computationally expensive but also less accurate considering the presence of boundary conditions in the simulation box.
A lot of effort in literature has been devoted to computing the long-range interactions efficiently, and widely studied
methods include lattice summation methods such as particle mesh Ewald (PME) \cite{darden1993particle,EPB+:JCP:1995}
and particle-particle particle mesh Ewald (PPPM) \cite{LDT+:MS:1994,Deserno98JCP}, and multipole type methods such as treecode \cite{BH:N:1986,DK:JCP:2000} and
fast multipole methods (FMM) \cite{GR:JCP:1987,GR:AN:1997,YBZ:JCP:2004}. These methods can reduce the operations per step
to $\cO(N\log N)$ or $\cO(N)$, and have gained big success in practice, but many problems remain to be solved as the prefactor
in the linear scaling can be large, or their implementation is nontrivial, or the scalability for parallel computing is not high.

The mainstream packages \cite{BBO+:JCC:1983,YTH+:JPCB:2006,HK+:JCTC:2008} for all-atom molecular dynamics simulations mostly use
Ewald-type lattice-summation algorithms which are originally proposed by Ewald \cite{ewald1921berechnung,de1980simulation}.
This type of methods split the Coulomb kernel
into a rapidly decaying function in the real space and a smooth function. % calculated in Fourier space.
The cutoff scheme is introduced for the first part in the real space. The smooth part is approximated by the Fourier series expansion.
The classical Ewald achieves an $\cO(N^{3/2})$ complexity to sum up all interactions.
When the cutoff radius is independent of $N$ and the Fourier series is accelerated by the fast Fourier transform (FFT) with an interpolation
to distribute charges on lattices, one obtains the PME method which achieves an $\cO(N\log N)$ complexity. The state-of-the-art development of
the Ewald-type algorithm includes an optimized choice of volumetric decomposition FFT scheme
for large systems on massively parallel supercomputers \cite{jung:JCC:19} and efficient methods for Coulomb interactions
without full periodicity \cite{girotto17,PanYiHu17PCCP}.

In this work, we propose a random batch Ewald (RBE) method of particle systems with Coulomb interactions
which enables an $\cO(N)$ Ewald method for fast molecular dynamics simulations. The ``random mini-batch'' idea,
namely using the sum over a small random set to approximate
some big summation, has its origin in the stochastic gradient descent (SGD) method \cite{robbins1951stochastic,bottou1998online}.
This type of ideas have been developed into different methods such as the stochastic gradient Langevin dynamics for Bayesian inference \cite{welling2011bayesian},
stochastic binary interaction algorithms for the mean field swarming dynamics \cite{albi2013}, the random batch method for interacting particle systems \cite{jin2020random},
and random-batch Monte Carlo simulations \cite{li2020Random}.  Though the specific implementations are different for different applications, these methods
are intrinsically Monte Carlo methods for computing the big summation involved in the dynamics, and the convergence can be obtained due to a time averaging effect \cite{jin2020random}, obeying the law of large numbers in time.

The RBE method uses the same idea of random minibatch. The new design being different from previous work is that the minibatch is built into the Ewald summation and sampled from the Fourier space. We take a cutoff radius in the real space such that the particles within the radius is of order one, and sample $p=\mathcal{O}(1)$ frequencies in the Fourier expansion of the smooth part of the Ewald splitting.
These $p$ frequencies are chosen into the minibatch for the force calculation.
The advantages of this approach are threefold. First, the short-range part of the force remains exact and thus the variance of the force
can be significantly reduced. Second, the short-range repulsive force due to the van der Waals interaction can be naturally introduced to avoid unphysical
configuration. Third and the most important, the importance sampling can be used in the Fourier space in building the minibatch such that the force variance can be further reduced. These strategies combined lead to a simple and efficient RBE method for molecular dynamics,
as shown in our numerical examples for calculating typical properties of electrolytes.

The rest of the paper is organized as follows. Section \ref{sec:preliminary} is devoted to an introduction to the setup and the classical Ewald summation,
which forms the basis of our method. In Section \ref{sec:rbewald}, we introduce the methodology of the RBE and give its detailed implementation.
We also provide some theoretic evidence on why the method works and can be efficient. In Section \ref{sec:numerical}, we test the RBE on
two typical problems to validate the method. Conclusions are given in Section \ref{sec:conclusion}.

\section{Overview of the Ewald summation} \label{sec:preliminary}

In this section, we introduce the setup and notations to be used later. Then, we give a brief review of the classical Ewald summation \cite{ewald1921berechnung,de1980simulation}.

To approximate electrostatic interactions between charges in an electrolyte of big extent, one often uses a simulation box with periodic boundary conditions (PBCs) \cite{frenkel2001understanding}
to mimic the bulk environment of the electrolyte. Without loss of generality, we consider a cubic box with the edge length $L$ so that the volume of the box is given by
$
V=L^3.
$
During the simulation, one calculates interactions of $N$ numerical particles (not necessarily the physical particles)
inside the box with positions $\bm{r}_i$ and charge $q_i$ ($1\le i\le N$)
satisfying the electroneutrality condition
\begin{gather}
\sum_{i=1}^N q_i=0.
\end{gather}
Then, one evolves the particles according to Newton's equations
\begin{gather}\label{eq:newton}
\begin{split}
&d\bm{r}_i=\bm{v}_i\,dt,\\
&m_i d\bm{v}_i=\bm{F}_i(\{\b{r}_i\})\,dt+d\bm{\eta}_i,
\end{split}
\end{gather}
where $d\b{\eta}_i$ represents the coupling with the environment (heat bath) (see Section \ref{subsec:thermostat}).

The forces are computed using $\bm{F}_i=-\nabla_{\b{r}_i}U$, where $U$ is the potential energy of the system.
Let $\bm{r}_{ij}:=\bm{r}_j-\bm{r}_i$ and $r_{ij}=|\bm{r}_{ij}|$ be the distance. The potential energy of the system due to Coulomb interactions with PBCs can be written as
\begin{gather}\label{eq:energy}
U=\frac{1}{2}\sum_{\bm{n}}{}'\sum_{i,j=1}^N q_iq_j \frac{1}{|\bm{r}_{ij}+\bm{n}L|},
\end{gather}
where $\bm{n}\in \mathbb{Z}^3$ ranges over the three-dimensional integer vectors and $\sum'$ is defined such that $\bm{n}=0$ is not included when $i=j$.
Due to the long-range nature of the Coulomb potential, this series converges conditionally. Hence, directly computing the interaction energy \eqref{eq:energy}
and the corresponding interaction forces using a cutoff approach is less accurate, and one shall introduce more advanced techniques to sum up the infinite series.

The idea of the classical Ewald summation is to separate the series into long-range smooth parts and short-range singular parts.
The conditional convergence due to the long-range, but smooth, parts
can be dealt with from the Fourier side.  To describe the details, one first introduces the error function
\[
\erf(x):=\frac{2}{\sqrt{\pi}}\int_0^x \exp(-u^2)du
\]
and the error complementary function is $\erfc(x):=1-\erf(x)$. Clearly, the Coulomb kernel can be written as
\[\frac{1}{r}=\frac{\erf(\sqrt{\alpha}r)}{r}+\frac{\erfc(\sqrt{\alpha}r)}{r}
\]
for any positive constant $\alpha$,
and the potential energy \eqref{eq:energy} can be decomposed as $U:=U_1+U_2$ with
\begin{gather}
U_1=\frac{1}{2}\sum_{\bm{n}}{}'\sum_{i,j}q_iq_j\frac{\erf(\sqrt{\alpha}|\bm{r}_{ij}+\bm{n}L|)}{|\bm{r}_{ij}+\bm{n}L|}, \\
~~U_2=\frac{1}{2}\sum_{\bm{n}}{}'\sum_{i,j}q_iq_j\frac{\erfc(\sqrt{\alpha}|\bm{r}_{ij}+\bm{n}L|)}{|\bm{r}_{ij}+\bm{n}L|}.
\end{gather}
The sum in $U_2$ now converges absolutely and rapidly, and one can truncate it to simplify the computation.
The sum in $U_1$ still converges conditionally in spite of the charge neutrality condition, but since the kernel is smooth,
the summation can be treated nicely in the Fourier domain. Define the Fourier transform:
\[
\widetilde{f}(\bm{k}):=\int_{\Omega}f(\bm{r}) e^{-i \bm{k}\cdot \bm{r}}d\bm{r},
\]
with $\bm{k}=2\pi\bm{m}/L$ and $\bm{m}\in \bbZ^3$. The inverse transform gives $f(\bm{r})=(1/V)\sum_{\bm{k}} \widetilde{f}(\bm{k})e^{i\bm{k}\cdot \bm{r}}$.
Then,  $U_1$ is expressed as (see \cite[Chap. 12]{frenkel2001understanding}),
\begin{gather}
U_1=\frac{2\pi}{V}\sum_{\bm{k}\neq 0}\frac{1}{k^2}|\rho(\bm{k})|^2
e^{-k^2/4\alpha}-\sqrt{\frac{\alpha}{\pi}}\sum_{i=1}^N q_i^2,
\end{gather}
where $k=|\bm{k}|$ and $\rho(\bm{k})$ is given by
\begin{gather}
\rho(\bm{k}):=\sum_{i=1}^N q_i e^{i\bm{k}\cdot\bm{r}_i},
\end{gather}
which can be viewed as the conjugate of the Fourier transform of the charge density. The divergent $\bm{k}=0$ term is
usually neglected in simulations to represent that the periodic system is embedded in a conducting medium which is essential for simulating ionic systems.

By inspection of the expressions above, we may take truncations in both the real and frequency domains. In particular, picking the real space and the reciprocal space cutoffs
\begin{gather}\label{eq:cutoffs}
r_c:= s/\sqrt{\alpha},~~ k_c: =2s\sqrt{\alpha},
\end{gather}
one then has (see \cite{kolafa1992cutoff})
\begin{multline}\label{eq:ewaldsumU}
U=\frac{2\pi}{V}\sum_{0<k\le k_c}\frac{1}{k^2}|\rho(\bm{k})|^2
e^{-k^2/4\alpha}-\sqrt{\frac{\alpha}{\pi}}\sum_{i=1}^N q_i^2+\\
\frac{1}{2}\sum_{|\bm{r}_{ij}+\bm{n}L|\le r_c}q_iq_j\frac{\erfc(\sqrt{\alpha}|\bm{r}_{ij}+\bm{n}L|)}{|\bm{r}_{ij}+\bm{n}L|}+
\mathcal{O}\left(\frac{Qe^{-s^2}}{s^2}(\frac{s}{\sqrt{\alpha}L^3})^{\frac{1}{2}}\right)
=:\widetilde{U}_1+\widetilde{U}_2+\mathcal{O}(\cdot),
\end{multline}
where $Q:=\sum_{i=1}^N q_i^2$, $\widetilde{U}_1$ is defined by the sum of the first two terms, and
$\widetilde{U}_2$ corresponds to the third term.
The density of particles in the real space $\rho_r$ and density of frequencies $\rho_f$ in the reciprocal space are given respectively by
\begin{gather}
\rho_r=\frac{N}{L^3},~\hbox{and}~\rho_f=\left(\frac{L}{2\pi}\right)^3.
\end{gather}
The number of interacting particles to be considered for a given particle is thus
\[
N_r:=\frac{4\pi}{3}r_c^3\rho_r=\frac{4\pi s^3 N}{3\sqrt{\alpha}^3L^3},
\]
 yielding total pairs $N_p=(4\pi/3) s^3 N^2/(\sqrt{\alpha}L)^3$.
The number of frequencies to be considered is  $N_f=(4\pi/3) k_c^3 \rho_f=(4/3\pi^2)(sL\sqrt{\alpha})^3$. The total work to compute $\widetilde{U}_1$ is thus
$T_f\sim N_f N$ since the computation of $\rho(\bm{k})$ needs $\mathcal{O}(N)$ operations.
The total work to compute $\widetilde{U}_2$ is  $T_r \sim N_p$.
In the usual Ewald summation, one needs to balance these two parts of works, thus  $N_p\sim N_f N$. Hence $\sqrt{\alpha}\sim N^{1/6}/L$ is chosen to balance the costs between
the real and frequency domains. This then yields the total number of pairs $N_p=\mathcal{O}(N^{3/2})$, and the number of frequencies to be considered is given by $N_f=\mathcal{O}(N^{1/2})$ so that the complexity in the frequency part is $T_f=\mathcal{O}(N^{3/2})$. Hence, the total complexity per iteration is  $\mathcal{O}(N^{3/2})$ for the energy computation.

The computation of force can be done directly using
\begin{multline}\label{eq:force}
\bm{F}_i=-\nabla_{\bm{r}_i}U=-\sum_{\bm{k}\neq 0}\frac{4\pi q_i \bm{k}}{V k^2}
e^{-k^2/(4\alpha)}\mathrm{Im}(e^{-i\bm{k}\cdot\bm{r}_i}\rho(\bm{k}))\\
-q_i\sum_{j,\bm{n}}{'} q_j G(|\bm{r}_{ij}+\bm{n}L|)\frac{\bm{r}_{ij}+\bm{n}L}{|\bm{r}_{ij}+\bm{n}L|}=:\bm{F}_{i,1}+\bm{F}_{i,2},
\end{multline}
where we recall $\bm{r}_{ij}=\bm{r}_j-\bm{r}_i$, pointing towards particle $j$, and
\[
G(r):=\frac{\erfc(\sqrt{\alpha}r)}{r^2}+\frac{2\sqrt{\alpha}e^{-\alpha r^2}}{\sqrt{ \pi}r}.
\]
Note that the force $\bm{F}_{i,1}$ is bounded for small $\bm{k}$. In fact, $k \ge  2\pi/L$, so $Vk$ is not small.

Again, we are going to take the truncations as shown in Eq. \eqref{eq:cutoffs}.
With the choice $\sqrt{\alpha}\sim N^{1/6}/L$, there are $N_f=\mathcal{O}(N^{1/2})$ frequencies to consider. Note the $\rho(\bm{k})$ computed can be used for all $i$, so the complexity for computing the forces $\bm{F}_{i,1}$ for all $i=1,\cdots, N$ is $\mathcal{O}(N^{3/2})$. Since there are $\mathcal{O}(N^{1/2})$ particles to consider for each $i$, the complexity for computing the forces $\bm{F}_{i,2}$ for all $i=1,\cdots, N$ is also $\mathcal{O}(N^{3/2})$. The total complexity per iteration is thus $\mathcal{O}(N^{3/2})$.

It is remarked that the PPPM  \cite{LDT+:MS:1994,Deserno98JCP} is a fast way to compute the Ewald sum using the FFT. The PPPM chooses parameter $\alpha$ such that $\sqrt{\alpha}\sim N^{1/3}/L$.
Using the cutoffs \eqref{eq:cutoffs}, the number of frequencies to be considered and the number of particles in real space for a given particle are given respectively by
\begin{gather}
N_f=\mathcal{O}(N),~~\hbox{and} ~N_r=\mathcal{O}(1).
\end{gather}
Hence, to compute the force, the complexity corresponding to the summation of all frequency components is $\mathcal{O}(N\cdot N_f)=\mathcal{O}(N^2)$ in the direct Ewald summation.
To speed up the summation in the Fourier space, one meshes the simulation box, assigns charges on the grid by interpolation and then takes advantage of the FFT to obtain $\rho(\bm{k})$ so that the potential can be computed with $\mathcal{O}(N\log N)$ complexity. The potential and forces at the particle locations are then obtained by further interpolation and some numerical difference schemes. Hence, the complexity per iteration is $O(N\log N)$ for the PPPM.

\section{The random batch Ewald}\label{sec:rbewald}

We now aim to develop the stochastic molecular dynamics using the idea of random mini-batch. The implementation of mini-batch
(i.e., finding suitable cheap unbiased stochastic approximation) depends on the specific applications. For interacting particle systems in \cite{jin2020random},
the strategy is the random grouping of particles.
By inspection of the Ewald summation (\eqref{eq:ewaldsumU} and \eqref{eq:force}), we found that $e^{-k^2/(4\alpha)}$ is summable so that
it can be normalized to form a probability distribution. Hence, this allows us to do the importance sampling in the Fourier space. This leads to a random batch strategy for the simulations of molecular dynamics.

\subsection{The algorithm}

Let us consider the factor $e^{-k^2/(4\alpha)}$ within the first term in Eq. \eqref{eq:force}.
Denote the sum of such factors by
\begin{gather}\label{eq:S}
S:=\sum_{\bm{k}\neq 0}e^{-k^2/(4\alpha)}=H^3-1,
\end{gather}
where
\begin{gather}
H:=\sum_{m\in \bbZ}e^{-\pi^2 m^2/(\alpha L^2)}
=\sqrt{\dfrac{\alpha L^2}{\pi}}\sum\limits_{m=-\infty}^{\infty}e^{-\alpha m^2L^2}, \label{psf}
\end{gather}
Here, $S$ is the sum for all three-dimensional vectors $\bm{k}$ except $0$. The number $H$ is the one for one-dimensional sum.
The second equality in Eq. \eqref{psf} is obtained by the Poisson summation formula \cite{Benedotto1997sampling,MR3289059}.
Eq. \eqref{psf} can then ben simply truncated at $m=\pm 1$ to obtain an approximation,
\[
H\approx\sqrt{\frac{\alpha L^2}{\pi}}(1+2e^{-\alpha L^2})
\]
using the rapid convergence of the series as typical setup in our simulations
holds $\alpha L^2 \gg 1$. One can improve the accuracy by using more terms if needed.
Then, we have the exact expression for the probability
\begin{gather}\label{eq:probexpression}
\mathscr{P}_{\bm{k}}:=S^{-1}e^{-k^2/(4\alpha)},
\end{gather}
which, with $\bm{k}\neq 0$, is a discrete Gaussian distribution and can be sampled efficiently as detailed below.

We apply the Metropolis-Hastings (MH) algorithm (see \cite{hastings1970monte} for details) to sample from the discrete distribution \begin{gather}\mathscr{P}(m)\sim H^{-1}e^{-(2\pi m/L)^2/4\alpha}.
\end{gather}
Doing this sampling procedure for three independent experiments will generate the components $k_i$ ($i=1,2, 3$) of $\bm{k}$ as $\bm{k}= 2\pi \bm{m}/L$.
The samples with $k_1=k_2=k_3=0$ will be discarded. In the MH procedure, the proposal $m^*$ is generated by first drawing
$x^*\sim \mathcal{N}(0, \alpha L^2/(2\pi^2))$, the normal distribution with mean zero and variance $\alpha L^2/(2\pi^2)$,
and one then sets $m^{\ast}=\mathrm{round}(x^{\ast})$, which is accepted with probability $q(m^{\ast}|m)$
in the MH algorithm, and clearly the probability is given by the following explicit expression,
\begin{equation}
\begin{split}
q(m^{\ast}|m)&=\int_{m^{\ast}-1/2}^{m^{\ast}+1/2}\sqrt{\dfrac{\pi}{\alpha L^2}}e^{-\pi^2 x^2/\alpha L^2}dx\\
&=\begin{cases}
\erf\left(\dfrac{1/2}{\sqrt{\alpha L^2/\pi^2}}\right)& m^{\ast}=0\\
\dfrac{1}{2}\left[\erf\left(\dfrac{|m^{\ast}|+1/2}{\sqrt{\alpha L^2/\pi^2}}\right) -\erf\left(\dfrac{|m^{\ast}|-1/2}{\sqrt{\alpha L^2/\pi^2}}\right)\right]& m^{\ast}\neq 0.
\end{cases}
\end{split}
\end{equation}
Since $\mathscr{P}(m^{\ast})\approx q(m^{\ast}|m)$, the acceptance rate is very high, which leads to small errors in this sampling procedure.
In practical implementation, one can precompute $q(m^*|m)\equiv \bar{q}(m^*)$ for a large enough range of $m^*$ values to speed up the sampling procedure.

We now consider the calculation of the forces in Eq. \eqref{eq:force} using the random mini-batch strategy. At each step, one picks a batch size $p$,
which is of $\mathcal{O}(1)$, and draws $p$ frequencies $\bm{k}_{\ell}$, $1\le \ell \le p$, i.i.d. from the discrete distribution $\mathscr{P}_{\bm{k}}$
by the MH sampling method above. The force $\bm{F}_{i,1}$ in \eqref{eq:force} is then approximated by the following random variable:
\begin{gather}\label{eq:rbmapprox}
\bm{F}_{i,1}\approx \bm{F}_{i,1}^*:=-\sum\limits_{\ell=1}^p \dfrac{S}{p}\dfrac{4\pi \bm{k}_\ell q_i}{V k_\ell^2}\mathrm{Im}(e^{-i\bm{k}_\ell\cdot\bm{r}_i}\rho(\bm{k}_\ell)).
\end{gather}
In the molecular dynamics simulations, we use this stochastic force $\bm{F}_{i,1}^*$ which is unbiased for the force calculation to replace $\bm{F}_{i,1}$.
The resulted molecular dynamics is a much cheaper version of the Ewald summation, and we call this stochastic method the Random Batch Ewald (RBE).

Of course, we need to update the $p$ samples after each time iteration. Suppose we have picked a step size $\Delta t$ and defined the time grid $t_n=n\Delta t$.  Then, we renew the batch of frequencies at each time grid point $t_n$. In real simulations, one will also add the van der Waals potential such as the Lennard-Jones potential
so that positive and negative charges will not merge. The force on each particle is then calculated by summing up the contributions
of real and Fourier parts, and the Lennard-Jones force (and other forces such as chemical bonds if any). Then, one
integrates Newton's equations \eqref{eq:newton} to obtain the position and velocity of the particle in the next time step.
Algorithm \ref{RBEalg} shows one possible such molecular dynamics method using the RBE
with some appropriate thermostat coupled to a heat bath (see Section \ref{subsec:thermostat} for discussions).

\begin{algorithm}[H]
	\caption{(Random-batch Ewald)}\label{RBEalg}
	\begin{algorithmic}[1]
		\State Choose $\alpha$, $r_c$ and $k_c$ (the cutoffs in real and Fourier spaces respectively), $\Delta t$, and batch size $p$.  Initialize the positions and velocities of charges $\bm{r}^0_i, \bm{v}^0_i$ for $1\le i\le N$.
                 \State Sample sufficient number of $\bm{k}\sim e^{-k^2/(4\alpha)},\, \bm{k}\neq 0$ by the MH procedure to form a set $\mathcal{K}$.
		\For {$n \text{ in } 1: N$}
		\State Integrate Newton's equations \eqref{eq:newton} for time $\Delta t$ with appropriate integration scheme and some appropriate thermostat. The Fourier parts of the Coulomb forces are computed using RBE force \eqref{eq:rbmapprox} with the $p$ frequencies chosen from $\mathcal{K}$ in order.
		\EndFor
	\end{algorithmic}
\end{algorithm}

In the case of the leapfrog scheme (equivalent to velocity-Verlet method) and the Andersen thermostat, the loop step in Algorithm \ref{RBEalg} is as follows.
\begin{enumerate}[(1)]
\item Choose $p$ frequencies from $\mathcal{K}$ without replacement; calculate real and Fourier parts of the electrostatic Coulomb force using RBE \eqref{eq:rbmapprox}, and other forces such as the Lennard-Jones forces.
\item Update the position and velocity of each particle using the following scheme for $n\ge 1$
	\[
	\begin{split}
	&\bm{v}_i^{n+1/2} = \bm{v}_i^{n-1/2}+\frac{1}{m_i}\bm{F}_i^{n} \times \Delta t,\\
	&\bm{r}_i^{n+1} = \bm{r}_i^n+\bm{v}_i^{n+1/2} \times \Delta t.
	\end{split}
	 \]
	(Here, $\bm{v}_i^{1/2}$ can be obtained via the  Euler scheme.)
\item  Update the velocity $\bm{v}_i^{n+1/2}$ of each particle with probability $\nu \Delta t$ by resampling $\b{v}_i$ from the normal distribution $\mathcal{N}(0, I_3T/m_i)$.
\end{enumerate}

We now analyze the complexity of the RBE method per time step.
Similar to the strategy in the PPPM, we may choose $\alpha$ such that the time cost in real space is cheap and the computation in the Fourier space is then accelerated.
Compared to the PPPM, the only difference is that the PPPM uses FFT and the RBE uses random mini-batch idea to speed up the computation in the Fourier space. Hence, we make the same choice,
\[
\sqrt{\alpha}\sim \frac{N^{1/3}}{L}=\rho_r^{1/3},
\]
which is inverse of the average distance between two numerical particles.
The complexity for the real space part is $\mathcal{O}(N\cdot N_r)=\mathcal{O}(N)$.
%The PME introduces FFT to sum up the frequency components to achieve $\mathcal{O}(N\log N)$ complexity.
Using the random batch approximation \eqref{eq:rbmapprox} which is a certain Monte Carlo method for approximating the force, the number of frequencies
to be considered is then reduced to
\begin{gather}
N_f=\mathcal{O}(p).
\end{gather}
If we choose {\it the same batch} of frequencies for all forces \eqref{eq:rbmapprox} (i.e., using the same
$\bm{k}_{\ell}$, $1\le \ell \le p$ for all $\bm{F}^*_{i,1}$) in the same time step, since the computed numbers $\rho(\bm{k}_{\ell})$ can be used for all particles, the complexity per iteration for the frequency part is reduced to $N_fN=\mathcal{O}(pN)$. This implies that the RBE method has linear complexity per time step if one chooses $p=\mathcal{O}(1)$.

\subsection{Consistency and stability}\label{subsec:analysis}

In this subsection, we provide some theoretic evidence for the
consistency and stability of the RBE algorithm in order to demonstrate
its validity.

According to Eq. \eqref{eq:probexpression}, we find that the long wave (low frequency) modes are more likely to
be chosen in the random approximation. Since the long wave modes are more important for the periodic effects,
this importance sampling strategy could be more effective compared with the uniform sampling across the modes considered.
This importance sampling strategy could also possibly reduce the variance so that the random method is more stable.
We now provide some theoretic evidence to explain why this method works.

We define the fluctuation in the random batch approximation for the Fourier part of the force on particle $i$ by,
\begin{gather}
\bm{\chi}_i:= \bm{F}_{i,1}^*-\bm{F}_{i,1}.
\end{gather}
The expectation and variance of the fluctuation can be obtained by direct calculation, which is given by Proposition \ref{pro:consistency}.
\begin{proposition}\label{pro:consistency}
The fluctuation in force $\bm{\chi}_i$ has zero expectation,
\begin{gather}
\E \bm{\chi}_i=0,
\end{gather}
and that the variance is,
\begin{gather}\label{eq:varianceforce}
\E |\bm{\chi}_i|^2 =\frac{1}{p}\left(\sum_{\bm{k}\neq 0}\frac{(4\pi q_i)^2S}{V^2 k^2}e^{-k^2/(4\alpha)}|\mathrm{Im}(e^{-i\bm{k}\cdot\bm{r}_i}\rho(\bm{k}))|^2-|\bm{F}_{i,1}|^2 \right).
\end{gather}
\end{proposition}

The first claim in Proposition \ref{pro:consistency} implies that the random approximation is consistent or unbiased,
\begin{gather}
\E \bm{F}_{i,1}^*=\bm{F}_{i,1},
\end{gather}
where $\E$ means expectation in probability theory (or the ensemble average in the physics community).
The second claim says that
\[
\E |\bm{\chi}_i|^2 \lesssim \frac{1}{p}\frac{S}{V}U_1= \frac{1}{p}\rho_r U_1.
\]
If the density $\rho_r=N/V$ is not very big, we expect our stochastic algorithm to work well.
Since for $k\gg \sqrt{\alpha}$, the factor $e^{-k^2/(4\alpha)}$ is very small and contributes
little to the variance in \eqref{eq:varianceforce}. Let us now consider the terms with $k\lesssim \sqrt{\alpha}$.
In the dilute solution regime where the Debye--H\"uckel (DH) theory (see \cite{levin2002electostatic},
and also Appendix \ref{app:linearpb}) is applicable, we expect that $|\mathrm{Im}(e^{-i\bm{k}\cdot\bm{r}_i}\rho(\bm{k}))|\approx 0$.
That means the variance is nearly zero. Of course, due to the deviation from the Debye--H\"uckel theory by thermal fluctuation,
this cannot be zero. We expect that $|\mathrm{Im}(e^{-i\bm{k}\cdot\bm{r}_i}\rho(\bm{k}))|$ does not change too much by the thermal fluctuation
for $k\ll a^{-1}$ where $a$ is the diameter of the ions (see Appendices \ref{app:linearpb}--\ref{app:forcevar}).
Clearly, if $\sqrt{\alpha}\ll a^{-1}$, the frequencies we consider then satisfy $k\ll a^{-1}$. We then can safely bound
\[
|\mathrm{Im}(e^{-i\bm{k}\cdot\bm{r}_i}\rho(\bm{k}))| \le C.
\]
In Appendix \ref{app:forcevar}, it is computed under this assumption that
\begin{gather}
\E |\bm{\chi}_i|^2 \lesssim \frac{1}{p}\rho_r^{4/3},
\end{gather}
which verifies that the variance of the random force is indeed controlled if the density is not big.

The following result indicates that random mini-batch methods can be valid for capturing the finite time dynamics
(we take the Langevin thermostat for illustration and see Section \ref{subsec:thermostat} for discussions).
\begin{theorem}
Let $(\bm{r}_i, \bm{v}_i)$ be the solutions to
\[
\begin{split}
&d\bm{r}_i=\bm{v}_i\,dt,\\
&m_i d\bm{v}_i=\left[\bm{F}_i(\{\bm{r}_j\})-\gamma \bm{v}_i\right]\,dt+\sqrt{2\gamma/\beta}d\bm{W}_i,
\end{split}
\]
where $\{\bm{W}_i\}$ are i.i.d. Wiener processes. Let $(\widetilde{\bm{r}}_i, \widetilde{\bm{v}}_i)$ be the solutions to
\[
\begin{split}
&d\widetilde{\bm{r}}_i=\widetilde{\bm{v}}_i\,dt,\\
&m_i d\widetilde{\bm{v}}_i=\left[\bm{F}_i(\{\widetilde{\bm{r}}_j\})+\bm{\chi}_i-\gamma \widetilde{\bm{v}}_i\right]\,dt+\sqrt{2\gamma/\beta}d\bm{W}_i,
\end{split}
\]
with the same initial values as $(\bm{r}_i, \bm{v}_i)$. Suppose that the masses $m_i$'s are bounded uniformly from above and below.
If the forces $\bm{F}_i$ are bounded and Lipschitz and $\E \bm{\chi}_i=0$, then for any $T>0$, there exists $C(T)>0$ such that
\[
\E\left[\frac{1}{N}\sum_i (|\bm{r}_i-\widetilde{\bm{r}}_i|^2
+|\bm{v}_i-\widetilde{\bm{v}}_i|^2) \right] \le C(N,T)\sqrt{\Lambda \Delta t},
\]
where $\Lambda$ is an upper bound for $\max_i (\E|\bm{\chi}_i|^2)$.
\end{theorem}
Similar proofs for interacting particle systems can be found in \cite{jin2020convergence,li2019stochastic,li2020Random}, and we omit the proof for the claims here.  The constant $C(N,T)$ can be made independent of $N$ in the mean field regime \cite{jin2020convergence}. Clearly, due to the assumption that $\bm{F}_i$ is bounded and Lipschitz, the claims above are not helpful for our problem. Anyhow, it can give us some insight how random batch type methods work.
Clearly, for a given configuration, a force computed using the RBE is a random approximation to the true force. A single-step evaluation of such random force definitely has no accuracy compared to the true force. The intuition why such methods work is that the effects of random forces accumulate in time. Since the random forces are unbiased, the random errors will roughly cancel out over time. This ``law of large numbers'' type mechanism in time then makes the random method work. The error bound above is the square root of  variance multiplied by $\Delta t$, which is the typical error bound given by central limit theorem.
Hence, our method is not aiming at computing the forces correctly for a fixed configuration. Instead, we attempt to obtain the evolution of the configurations and the equilibrium distribution with an acceptable error control. We use the RBE method only to speed up MD simulations and obtain configurations, and then use these configurations to compute the true energies, stress tensor (and pressure) using their definitions, without random batch approximation.

A question that may arise is whether one should wait for too many iterations before the ``law of large number'' mechanism takes effects to capture the long time properties (i.e. whether the random batch type methods will delay the mixing time for the equilibria too much). In \cite{jin2020convergence,li2020Random}, it has been shown that when some external confining fields are present, the mixing time for convergence to the global equilibrium with random batch is roughly the same as the one without random batch, as the error controls are uniform in time. When there are no helping external fields such as the cases we are considering here in a periodic box, whether random batch will delay the convergence to the thermal equilibrium is still a theoretically open question.  However, when heat bath is present, if the number of particles or modes is statistically large so that a few of them can capture the significant statistical properties, the few chosen representatives may give the correct statistical properties and the random batch methods may capture the correct macroscopic quantities without looping for too many iterations. Hence, we believe the RBE method can capture the long time statistical properties for the many-body systems in contact with heat bath, without increasing the iterations of simulation too much.

As we have seen, the variance of the fluctuation is always multiplied by the step size $\Delta t$ in the error estimates: $\sqrt{\Lambda \Delta t}$ for the error of trajectories or $\Lambda \Delta t$ for the distributions (see \cite{jin2020convergence} for the weak error estimates regarding first order systems). Hence, the variance somehow measures the stability of the random methods and the boundedness of $\E|\bm{\chi}_i|^2$ is important for the convergence of the random algorithms.
Though the variance is controlled for the RBE, rigorous proof for this method is challenging as the field $\bm{F}_{i,2}$ is singular. Building in van der Waals potential into the system can prevent the particles getting too close so the singularity of $\bm{F}_{i,2}$ might not be seen, but the rigorous justification of convergence could still be very hard. We will leave the rigorous mathematical analysis for future exploration.

\subsection{Discussion on the thermostats}\label{subsec:thermostat}

To couple with the heat bath so that the temperature is preserved near the desired value, typical ways
include the Andersen thermostat and the Langevin thermostat. Another thermostat used in molecular
dynamics in a deterministic approach is the famous Nos\'e-Hoover thermostat \cite[Chap. 6]{frenkel2001understanding}.

In the Andersen thermostat, one assumes the collision frequency between the particle and
the heat bath is $\nu$. Then, the time between two collisions for a particular particle satisfies the exponential distribution.
Hence, the probability that a particle does not collide with the heat bath during $[t_n, t_n+\Delta t)$ is
$1-\exp(-\nu \Delta t)\approx \nu \Delta t$. Hence, in the Andersen thermostat, at each time grid $t_n$, for each particle
one resets the velocity with probability $\nu\Delta t$. The new velocity is sampled from the Maxwell distribution with
temperature $T$. In Eq. \eqref{eq:newton}, $d\b{\eta}_i$ corresponds to such resetting noise to model the collision.
It is this new velocity that guarantees the correct temperature. Of course, the larger $\nu$ is,
the better the temperature can be kept around the desired value. However, too large $\nu$ value will bring
some unphysical effects \cite{nose1984molecular,hoover1985canonical,frenkel2001understanding}.

In the Langevin thermostat, the interaction with heat bath is added into the equation directly:
\[
m_id\bm{v}_i=(\bm{F}_i-\gamma\bm{v}_i)dt+\sqrt{2\gamma/\beta}\,d\bm{W}_i.
\]
In other words, in \eqref{eq:newton}, $d\b{\eta}_i=-\gamma\bm{v}_idt+\sqrt{2\gamma/\beta}\,d\bm{W}_i$.
The term $-\gamma \bm{v}_idt$ is the friction and $\sqrt{2\gamma/\beta}\,d\bm{W}_i$
is the thermal noise or the fluctuation, both arising from the collision with the heat bath.
The fluctuation-dissipation relation requires the strength of the noise to be $\sqrt{2\gamma/\beta}$
so that the system can tend to the correct temperature $T=\beta^{-1}$ (the Boltzmann constant $k_B$
is taken to be $1$ for the reduced units). As in the Andersen thermostat, increasing $\gamma$ can keep the temperature
of the system around $T$ better. However, since the temperature enters in through the dynamics, the Andersen thermostat
seems to behave better for the temperature control than the Langevin dynamics.

The Nos\'e-Hoover thermostat uses a Hamiltonian for an extended system of $N$ particles plus an additional coordinate $s$ (\cite{nose1984molecular,hoover1985canonical}):
\[
\mathcal{H}_{\mathrm{NH}}=\sum_{i=1}^N\frac{|\bm{p}_i|^2}{2m_i s^2}+U(\{\bm{r}_i\})
+\frac{p_s^2}{2Q}+L\frac{\ln s}{\beta}.
\]
Here, $\bm{p}_i=m_i \bm{v}_i$ is the momentum of the $i$th particle.
The microcanonical ensemble corresponding to this Hamiltonian reduces to the canonical ensemble for the real variables $\bm{p}_i'=\bm{p}_i/s$. Hence, one may run the following deterministic ODEs, which are the Hamilton ODEs under $\mathcal{H}_{\mathrm{NH}}$ in terms of the so-called real variables,
\[
\begin{split}
&\dot{\bm{r}}_i=\frac{\bm{p}_i}{m_i},\\
& \dot{\bm{p}}_i=-\nabla_{\bm{r}_i}U-\xi \bm{p}_i,\\
& \dot{\xi}=\frac{1}{Q}\left(\sum_{i=1}\frac{|\bm{p}_i|^2}{m_i}-\frac{3N}{\beta}\right).
\end{split}
\]
The time average of the desired quantities such as those in \eqref{fluc} will be the correct canonical ensemble average. As one can see, when the temperature of the system, defined by $\sum_{i=1}^N m_i |\bm{v}_i|^2/(3Nk_B)$, is different from $T$,
the extra term $-\xi \bm{p}_i$  will drive the system back to temperature $T$, and thus it may give better behaviors for controlling the temperature.

As we have seen, the random batch approaches will bring in extra variance term. Hence, there is numerical heating effect that increases the temperature by $\Lambda \Delta t$.
Due to this reason, the RBE is not suitable for long time simulation under NVE ensemble if without an appropriate conservation scheme for time integration,
but it should be good for NVT (as we do in this paper) and other simulations with thermostats.
To reduce this artificial temperature, one may on one hand reduce $\Lambda$ by using bigger
batch size or carefully designed importance samplings. The RBE proposed in this work is an importance sampling approach. Also, one may mimick the simulated annealing idea \cite{MR904050,MR995752,hwang1990large} to decrease $\Delta t$, which has also been used
in the stochastic gradient Langevin dynamics.
On the other hand, using suitable thermostat may drive the temperature back to $T$
better. In principle, the Nos\'e-Hoover is the most effective for preserving temperature. If the frequency $\nu$ is chosen suitably in the Andersen thermostat, the temperature can be preserved well too.  As we see below in Section \ref{sec:numerical}, if $\nu$ and batch size $p$ are slightly bigger and the simulated annealing approach is used, the Andersen thermostat is already enough for the numerical examples we consider. Due to the simplicity, we adopt the Andersen thermostat in this paper to illustrate that the RBE works, while leaving the Nos\'e-Hoover thermostat for our future development for large systems.

\section{Application examples}\label{sec:numerical}

In this section, we consider two typical application examples to validate the accuracy and efficiency of the proposed method.
The first example is the charge distribution in terms of charge-charge correlation functions  in an electrolyte solution with the primitive model where
the DH theory can be used to provide a theoretical prediction. The second example is a much harder example with
many different species of ions (including a macroion) where charge reversal phenomenon for electric
double layer near the surface of a colloidal particle is studied.
Both examples indicate that the proposed method is effective and efficient.
The calculations are performed in a Linux system with Intel Xeon Scalable Cascade Lake 6248@2.5GHz, 1 CPU core and 4 GB memory.

\subsection{Charge correlation functions in electrolyte}

In this example, we consider a pure electrolyte monovalent binary ions. The primitive model of the electrolyte
is employed, which describes the solvent as mobile ions of uniform sizes embedded in a medium of constant permittivity
under a given temperature. The total potential energy of the system is composed of the Coulomb interactions and the
short-range van der Waals interaction. The latter is modeled by the shifted Lennard-Jones potential expressed as:
\begin{eqnarray}\label{eq:Ljpotential}
V_{\mathrm{LJ}}(r)=
\begin{cases}
4\epsilon\left[\big(\dfrac{\sigma}{r-r_{\mathrm{off}}}\big)^{12}-\big(\dfrac{\sigma}{r-r_{\mathrm{off}}}\big)^6\right]
+V_{\mathrm{shift}},\qquad &\text{if}\quad r-r_{\mathrm{off}}<R_{c}\\
0, &\text{otherwise},
\end{cases}
\end{eqnarray}
where $r_{\mathrm{off}}=(d_i+d_j)/2-\sigma$ and $d_i$ and $d_j$ are the diameters of two particles respectively, and $\sigma$ is a positive constant.
$V_{\mathrm{shift}}$ is taken such that the potential becomes zero when $r-r_{\mathrm{off}}=R_{c}$.

All the quantities are provided in reduced units (see \cite[sec. 3.2]{frenkel2001understanding}). The diameter of each ion is chosen as $d_i\equiv 0.2$,
the reduced temperature is $T=1.0$ and the reduced dielectric constant is $\varepsilon=1/4\pi$ so that the electric potential of a charge $q$ is given by $\phi(r)=q/r$.
For the Lennard-Jones potential in this example, we choose the parameters as $\sigma=0.2$, $R_{c}=4.0$ and $\epsilon=1$.
We fix the particle density to be constant $N/L^3=0.3$. Correspondingly, the inverse Debye length in the Debye--H\"uckel theory (see Appendix \ref{app:linearpb}) is $\kappa\approx1.9416$.
We run molecular dynamics simulations to prepare the configuration samples and by taking average of these samples to
obtain the charge distribution in terms of charge-charge correlation functions,
\begin{gather}
\rho(r)=g_{++}(r)-g_{+-}(r),
\end{gather}
where $g_{++}$ and $g_{+-}$ are cation-cation and cation-anion pair correlation functions between ions.
By the Debye-H\"uckel theory, the radial distribution of net charge satisfies the following linear relation,
\begin{gather*}
\ln(r|\rho(r)|)\approx-1.9416r-1.1437,~~r\gg 0.2.
\end{gather*}
Here, due to the setting of the Lennard-Jones potential, we roughly have the parameter $a$ in Appendix \ref{app:linearpb} as
\[
a=\frac{1}{2}(d_1+d_2)-\sigma+\sigma=0.2,
\]
and the formula above should be accurate for $r\gg a$.

\begin{table}[ht]
	\centering
	\begin{tabular}{|c|c|c|c|c|c|}
		\hline
		&$\alpha$&$r_c$&$n_c$& steps & Time (s)\\\hline
		Ewald  &0.12&8.0&7&1e6& 6067 \\\hline
		PPPM  &0.55&4.0&15 &1e6 & 3120 \\\hline
		RBE  &0.55&4.0&$p=10$ &1e6 & 1267 \\\hline
	\end{tabular}
	\caption{Parameters and computational time for the Ewald, PPPM and RBE results with $N=300$. The RBE samples from all frequencies and it shows $p$ value in the $n_c$ column.}
	\label{tabl:iterationtSALT}
\end{table}

\begin{figure}[ht]	
	\centering
	\includegraphics[width=0.85\textwidth]{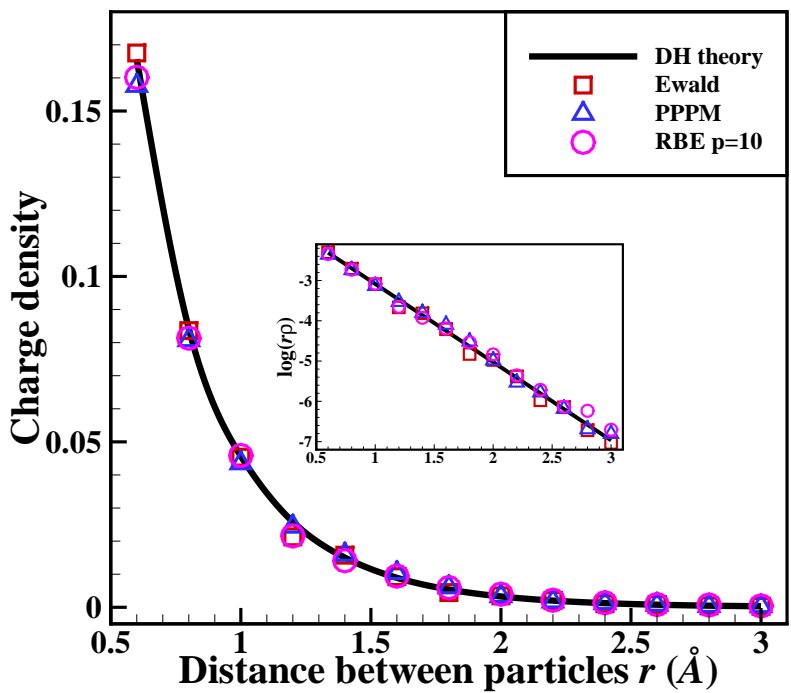}
	\caption{Charge density predicted by simulations using the RBE, classical Ewald  and PPPM methods for $L=10$ and $N=300$. }
	\label{fig:errcompare}
\end{figure}

In the first numerical experiment, we take the length of the periodic box to be $L=10.0$, and the number of monovalent ions $N=300$ so
that $n_{+1}=n_{-1}=150$. The Andersen thermostat is adopted with frequency $\nu=3$.
The parameters are chosen as in Table \ref{tabl:iterationtSALT}, where the $n_c$ column for RBE lists batch size $p$ as no frequency cutoff is introduced.
The parameters are chosen so that the estimated relative force errors for the Ewald method and PPPM are about $10^{-4}$ by \cite{kolafa1992cutoff} (the parameters are set automatically in LAMMPS software).
As discussed already, we choose the same $\alpha$ value for the RBE as that in the PPPM.
The batch size $p$ in the RBE is chosen through a convergence test and $p=10$ gives comparable results already. The results by the RBE, classical Ewald and PPPM methods
in comparison with those predicted by the DH theory are shown in Fig. \ref{fig:errcompare}.
It shows that the error by the RBE is comparable to those by the Ewald and PPPM methods. As Table \ref{tabl:iterationtSALT} indicates,
the computational time of the RBE is about $1/5$ of that for the Ewald method, $2/5$ of that for the PPPM method to achieve comparable results in spite that the system size is not very large.

\begin{figure}[ht]	
	\centering
	\includegraphics[width=0.49\textwidth]{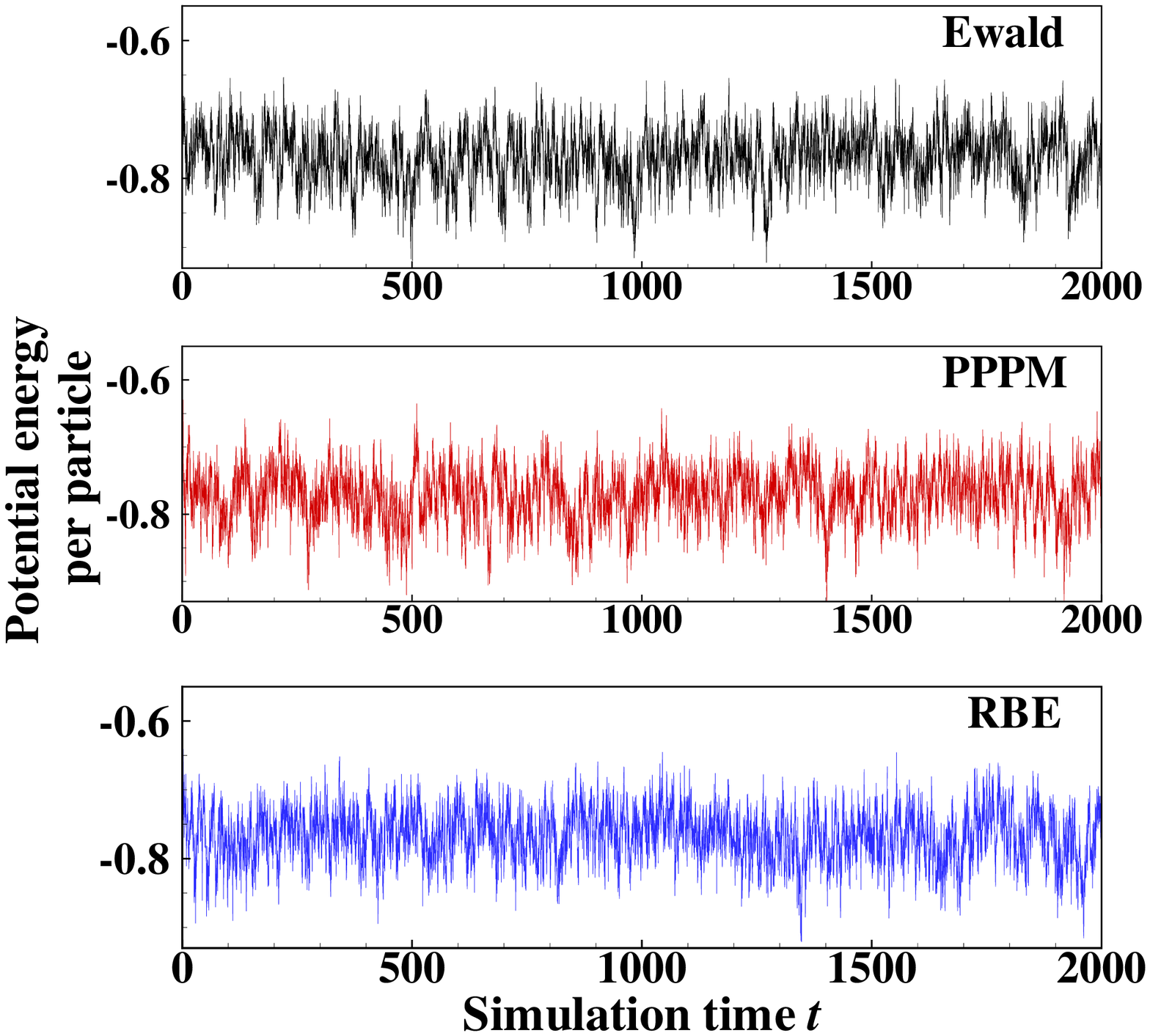}
	\includegraphics[width=0.49\textwidth]{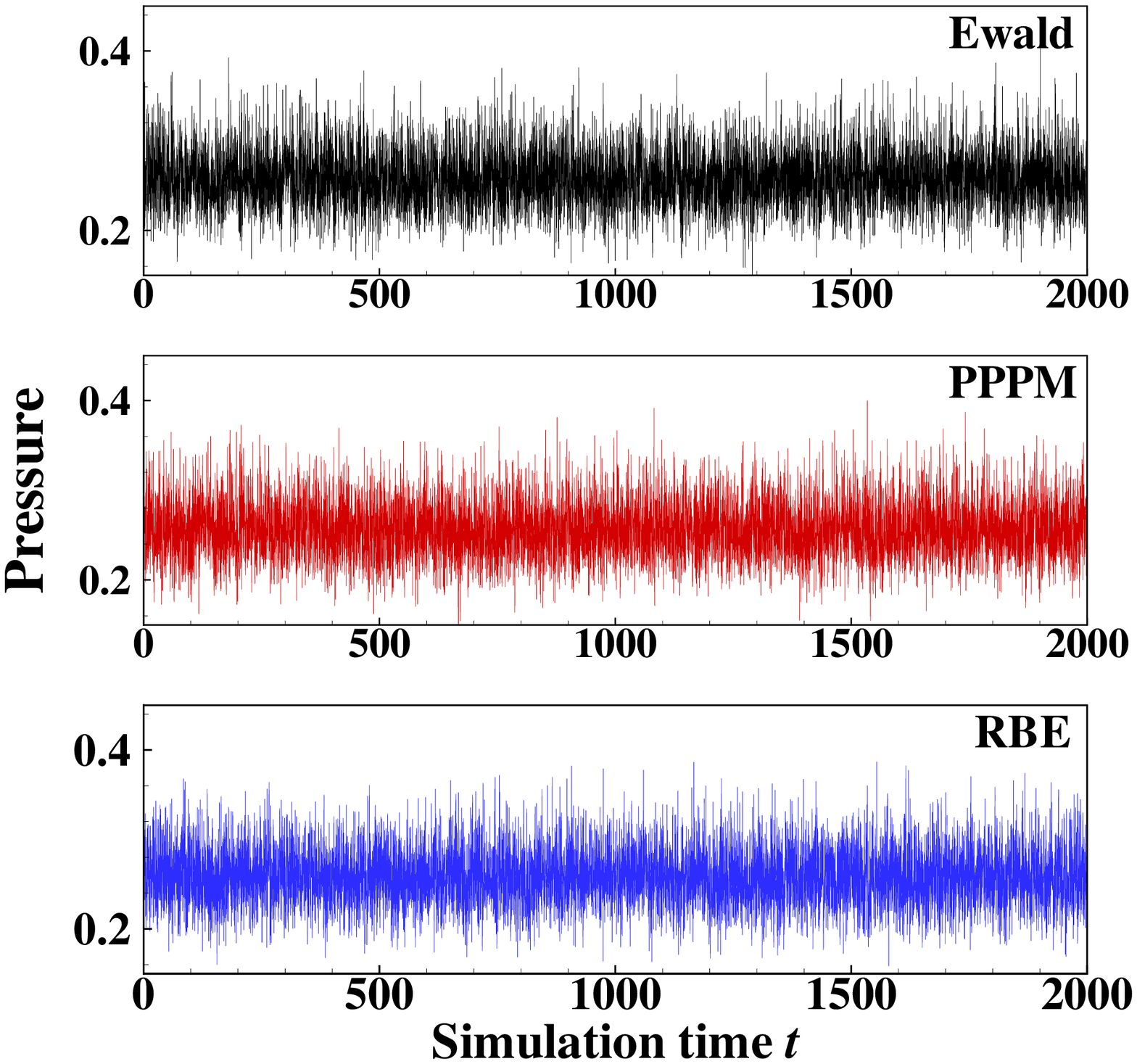}
	\caption{Potential energy per particle and pressure by simulation time using the RBE, classical Ewald and PPPM methods for $L=10$ and $N=300$. }
	\label{fig:energypressure}
\end{figure}

We also compute fluctuations of the potential energy (per particle) $E_\mathrm{pot}$ and pressure $P$ for the three methods to validate the correctness of the configurations.
These two quantities are defined by,
\begin{equation}\label{fluc}
\begin{split}
&E_\mathrm{pot}=\frac{1}{N}(E_\mathrm{Coul}+E_\mathrm{LJ}),\\
&P=\frac{2}{3V}\left(\sum_{i=1}^N\frac{1}{2}m_i \bm{v}_i^2 -\mathrm{vir}\right),
\end{split}
\end{equation}
where $\mathrm{vir}=(1/2)\sum_{i<j}\bm{r}_{ij}\cdot \bm{F}_{ij}$ is the virial
and $E_\mathrm{Coul}$ is calculated as \eqref{eq:ewaldsumU}. Pressure is calculated by using Clausius virial theorem with kinetic energy and virial tensor. We recall $\bm{r}_{ij}=\bm{r}_j-\bm{r}_i$, and $\bm{F}_{ij}$ is the force of particle $j$ acting on particle $i$. The potential energy per particle and the average pressure are calculated in LAMMPS using the virial formula \cite{brown1995general}.
In Fig. \ref{fig:energypressure}, the data of every $100$ time steps are plotted for the time up to $t=2000$.
We calculate the average data of these quantities, $\overline{E}_\mathrm{pot}$ and $\overline{P}$. The relative errors of
the RBE compared to the PPPM are both less than $1\%$.

\begin{figure}[ht]	
	\centering
	\includegraphics[width=0.85\textwidth]{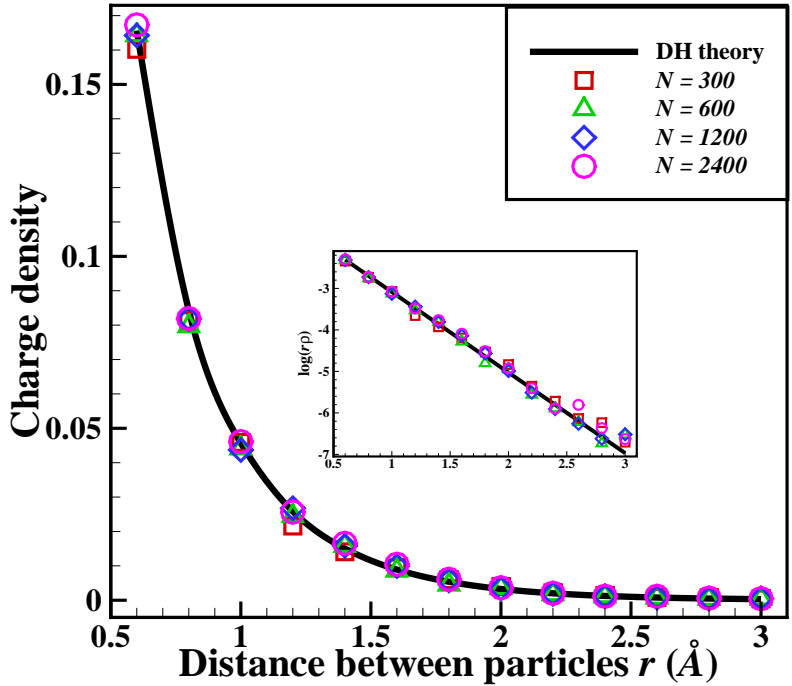}
	\caption{Charge density simulated by the RBE method with increasing system sizes for constant density. The batch size is $p=10$ for all $N$.}
	\label{fig:differentN}
\end{figure}

We increase the size of the system while keeping $\rho_r=N/L^3=0.3$ constant to measure the accuracy as well as
the computational time.  In particular, we choose $N=300, 600, 1200$ and $2400$,
respectively, and the length $L$ is computed correspondingly. In Fig. \ref{fig:differentN}, we show the simulation results
for the charge distribution with $p=10$ for the RBE method.
Clearly, the simulation results of the RBE method still agree well with the DH theory for larger $r$.
Particularly, in the embedded subplot we can observe that the linear relation holds up to the error tolerance $e^{-7}\approx 9.1\times 10^{-4}$,
which confirms the accuracy of the RBE method.

\begin{table}[ht]
	\centering
	\begin{tabular}{|c|c|c|c|}
		\hline
		&$\alpha$ & $r_c$&$n_c$\\\hline
		$N=100$  &0.55&4.0&10\\\hline
		$N=300$  &0.55&4.0&10\\\hline
		$N=1000$  &1.1&3.0&13\\\hline
		$N=4000$  &2.5&2.0&18\\\hline
	\end{tabular}
	\caption{Parameters for PPPM and RBE, where the RBE method does not have $n_c$. }
	\label{tabl:refpara1}
\end{table}
Next, we compute the relative accuracy of the potential energy for the RBE method against the PPPM for different densities. In particular, we fix $L=10$ and consider $N=100, 300, 1000$ and $4000$, respectively (correspondingly, $\rho_r=0.1, 0.3, 1.0$ and $4.0$). The parameters used in the calculations are shown in Table \ref{tabl:refpara1}. Note that the RBE method does not have $n_c$ parameter and instead we choose the batch size $p=10,20,50$ and $100$, respectively. The time step is again $\Delta t=0.002$. The potential energies are computed using $10^4$ configurations after equilibrium, sampled every $100$ steps.  The results are shown in Table \ref{tabl:relaerr}. Clearly, if we increase the density, we need to use larger batch size to get acceptable accuracy. The RBE with fixed batch size $p$ will have bigger error if the density is increased. Anyhow, even when $\rho_r=4.0$, using $p=100$ seems enough to get acceptable results.

\begin{table}[ht]
	\centering
	\begin{tabular}{|c|c|c|c|c|}
		\hline
	~~$\rho_r$~~	&$p=10$&$p=20$&$p=50$&$p=100$\\\hline
		$0.1$&$0.15\%$&$0.13\%$&$0.13\%$&$0.08\%$\\\hline
		$0.3$&$0.10\%$&$0.08\%$&$0.04\%$&$0.09\%$\\\hline
		$1.0$&$0.66\%$&$0.18\%$&$0.11\%$&$0.04\%$\\\hline
		$4.0$&$7.83\%$&$2.38\%$&$0.71\%$&$0.31\%$\\\hline
	\end{tabular}
	\caption{Relative error of potential energy for the RBE method against the PPPM method with different densities and batch sizes.}
	\label{tabl:relaerr}
\end{table}

\begin{figure}[ht]	
	\centering
	\includegraphics[width=0.85\textwidth]{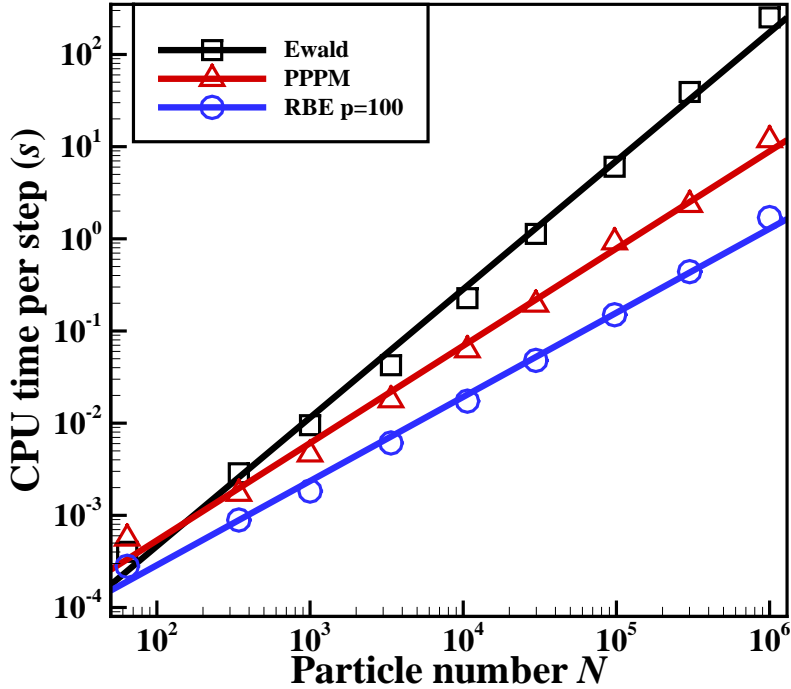}
	\caption{CPU time per step for the classical Ewald and RBE methods with increasing $N$.}
	\label{fig:N-t}
\end{figure}

Lastly, we compare the efficiency for the classical Ewald, PPPM and RBE methods.
In Fig. \ref{fig:N-t} the computational times for the three methods are shown for system size up to
$N=10^6$, where the solid lines present the linear fitting of the data in log-log scale. The
results agree with the fact that the complexity per time step for the Ewald summation is of
$\mathcal{O}(N^{3/2})$, while the complexity per time step for the RBE is only of $\mathcal{O}(N)$) and the complexity
per time step for the PPPM method is a little larger than $\mathcal{O}(N)$. The cost of the RBE is
small even when one chooses batch size $p = 100$. The RBE has the best efficiency over
a whole range of particle numbers, clearly demonstrating the attractive performance of
the algorithm. We remark that a systematic study of the efficiency of the method will be
performed in our next work for large-scale all-atom systems, in particular, the comparison with
the performance of the PPPM.

\subsection{Charge inversion in salty environment}

When a highly charged colloid in a solution that contains multivalent counterions, its charge can become overcompensated
due to the strong ion correlation between counterions, leading to the charge inversion (or charge reversal) phenomenon.
The many-body phenomenon has attracted much attention
in the past decades from both experimental \cite{BZH+:PRL:2004,PBV+:PRL:2006}, theoretical and simulation studies
\cite{GNS:RMP:2002,BKN+:PR:2005,MHK:CPC:2002,WM:JCP:2010,lenz2008simulation,GXX:JCP:2012}, since the charge inversion
implies that the effective charge of the colloid-microion complex is
abnormally inverted, opposite to the common intuition of understanding from the traditional Poisson-Boltzmann theory.

We follow the setup of Lenz and Holm \cite{lenz2008simulation} and consider
a highly charged colloid in a solution of asymmetric 3:1 salt
with additional 1:1 salt. The colloid has a spherical geometry of diameter  $d_0=100\Ai$
with a point charge $Q_0=-300e_0$ at its center. Here $e_0=1.6\times 10^{-19}C$ is the elementary charge.
It is placed  at the middle of a cubic box with the PBC. The side length of the periodic box is set
to $L = 225.8\Ai$ (the volume corresponds to a spherical cell of radius $R_0 = 140\Ai$).
Initially, a total of 200 trivalent counterions,  $300+n_{\mathrm{salt}}$ monovalent coions and
$n_{\mathrm{salt}}$ monovalent counterions are randomly distributed within the box. These
ions have uniform size with a diameter of $4\Ai$. Clearly, the system satisfies the
charge neutrality. The trivalent counterions correspond to the concentration of
$c_{(+3)}=30 ~\mathrm{mM}$ (i.e. milli mole per liter).

In this example,we implement the methods by the self-written molecular dynamics code with C++.
We focus on the accuracy comparison and investigate if or not the RBE can get the correct results
with less effort for this relatively complicated many-body phenomenon. Due to the strong charge of the colloid, we take
the classical Ewald results as the reference solution.
In the simulations, we consider two concentrations for the additional 1:1 salt, i.e.,
$c_{\mathrm{salt}}=0\,\mathrm{mM}$ and $c_{\mathrm{salt}}=196\,\mathrm{mM}$, where the latter case corresponds to the number of particles $n_{\mathrm{salt}}=1300$
for each ionic species.  The temperature is set to the room temperature $T_*=298K$, and the Bjerrum length $\ell_B$ is determined by
$\ell_B=e_0^2/(4\pi \varepsilon_0 \varepsilon_r k_BT_*)$,
where $\varepsilon_r=78.5$ is the relative dielectric constant of water and $\varepsilon_0$ is the vacuum permittivity,
resulting in $\ell_B=7.1\Ai$. The van der Waals interaction is again taken to be part of the Lennard-Jones potential \eqref{eq:Ljpotential}, where $R_{c}=2^{1/6} \Ai$, $\sigma=1 \Ai$ and $\epsilon= 1 k_BT_*$. Note that $r_{\mathrm{off}}$ is different for different pairs, e.g., $r_{\mathrm{off}}=52 \Ai$ between the colloid and an microion, and $r_{\mathrm{off}}=4 \Ai$ between microions.

To do simulations, we scale all lengths by $L_*=1 \Ai$, temperature by $T_*=298K$, and masses by $m_*$, the mass
of ions which are assumed equal. Then, other quantities can be scaled correspondingly: the energy is scaled by $k_BT_*$, the velocity by $(k_BT_*/m)^{1/2}$, and time by $L_*(m_*/k_BT_*)^{1/2}$, etc.. Consequently, in these reduced units, the room temperature becomes $T=1$, and the Coulomb interaction between two point charges $i$ and $j$ is given by $U_{ij}=\ell_B q_iq_j/r_{ij}$
where $\ell_B$ is the scaled Bjerrum length with value $7.1$.
After we computed the forces using formulas in Section \ref{sec:preliminary} or in Section \ref{sec:rbewald}, we should multiply the results with $\ell_B=7.1$
to get the Coulomb forces for this example.

The molecular dynamics simulations are all performed with the Andersen thermostat, with $1e5$ steps for the burn-in phase and $6e5$ steps for the sampling to compute ensemble averages. In the burn-in phase, we choose time step
\[
\Delta t_n=0.01/\ln(1+n),
\]
where $n$ is the number of time steps motivated by the simulated annealing mentioned above. In the sampling phase, we choose $\Delta t=0.002$.
The collision frequency $\nu=10$ for $c_{\mathrm{salt}}=0~\mathrm{mM}$, and $\nu=1$ for $c_{\mathrm{salt}}=196~\mathrm{mM}$. The reason to use smaller frequency for $c_{\mathrm{salt}}=196~\mathrm{mM}$ is to decrease the artificial diffusion effect introduced by the Andersen thermostat.

\begin{figure}[ht]	
	\centering
	\includegraphics[width=0.495\textwidth]{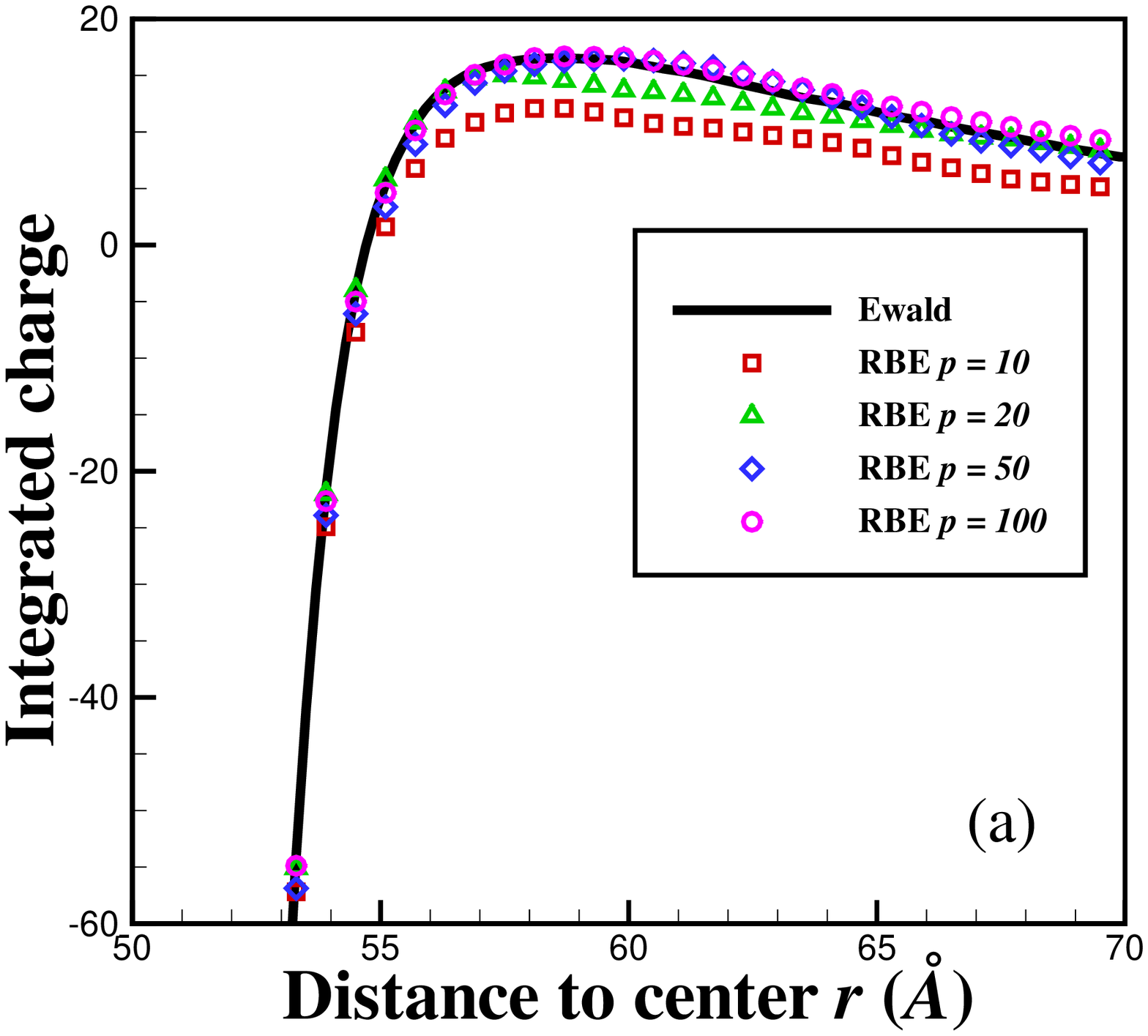}
	\includegraphics[width=0.495\textwidth]{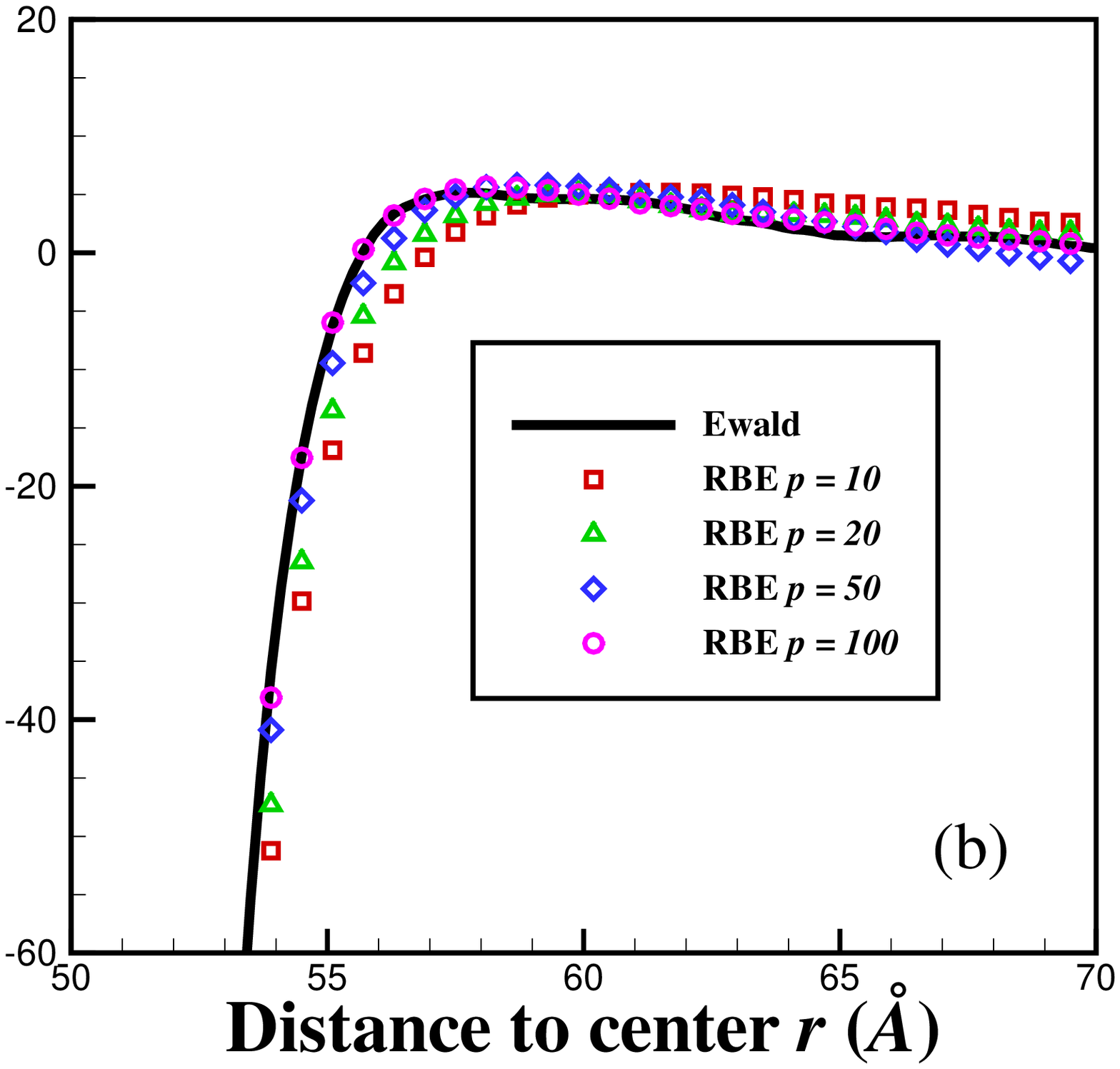}
	\caption{Integrated charge against the distance $r$ from the colloid center when $c_{\mathrm{salt}}=0~\mathrm{mM}$ (left) and $196~\mathrm{mM}$ (right): comparison of the Ewald and RBE methods. }
	\label{fig:Intcharge}
\end{figure}

\begin{table}[ht]
	\centering
	\begin{tabular}{|c|c|c|c|c|}
		\hline
		&$\alpha$&$r_c$&$n_c$& Time (s)\\\hline
		Ewald ($c=0$ mM)   &0.0014&90.0&8.7&16698 \\\hline
		RBE ($c=0$ mM)     &0.0072&40.0&  $p=100$  &1167 \\\hline
		Ewald ($c=196$ mM) &0.0014&90.0&8.7&137217 \\\hline
		RBE ($c=196$ mM)   &0.0072&40.0&  $p=100$  &15258 \\\hline
	\end{tabular}
	\caption{Computational time per 1e5 simulation steps. The RBE samples from all frequencies and it shows $p$ values in the $n_c$ column.}
	\label{tabl:iterationtCOLLIOD}
\end{table}

The settings and running time are
shown in Table \ref{tabl:iterationtCOLLIOD}. Clearly, the time consumption of the RBE method is much less (about $1/10$ of that for the Ewald method), so the proposed RBE is efficient.
The integrated charge distribution, the total charge within the radial direction distance, against the distance $r$
from the colloidal center is plotted in Fig. \ref{fig:Intcharge}. Regarding the effectiveness, as can be seen from the figure, the RBE can capture
the charge reversal phenomenon correctly and obtain acceptable simulation results.
As discussed in Proposition \ref{pro:consistency} and Section \ref{subsec:thermostat}, the force approximation is unbiased, but the randomness results
in positive variance leads to numerical heating and systematic error for the equilibrium distribution.  As can be seen in Fig. \ref{fig:Intcharge},
the overcharging effect is weakened for small batches due to this numerical heating. The RBE method converges after $p\gtrsim 100$ and this systematic error
is negligible for the system considered here. The inverted charge (maximum of the curve) by the RBE when $p\gtrsim 100$ is in agreement with the Ewald summation and the literature result \cite{lenz2008simulation}. This agrees with the discussion above in Sections \ref{subsec:analysis}--\ref{subsec:thermostat}. Since $\nu$ is smaller for $c_{\mathrm{salt}}=196~\mathrm{mM}$, the ability of temperature control is reduced and the numerical heating is more obvious for small $p$ values (like $p=20$).
To resolve this, one may consider subtracting the effective temperature due to the random batch from the desired $T$ value,
or using better thermostat such as the Nos\'e-Hoover thermostat. These issues will be explored in our subsequent work.

\begin{figure}[ht]	
	\centering
	\includegraphics[width=0.85\textwidth]{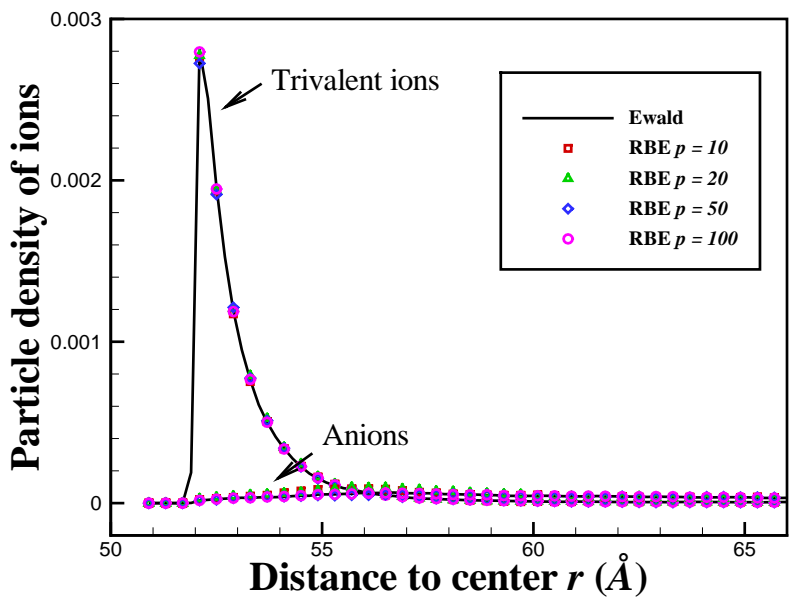}
	\caption{Contribution of different ion types to particle density $\rho$ when $c_{\mathrm{salt}}=0~\mathrm{mM}$.}
	\label{fig:chargeden0}
\end{figure}

\begin{figure}[ht]	
	\centering
	\includegraphics[width=0.85\textwidth]{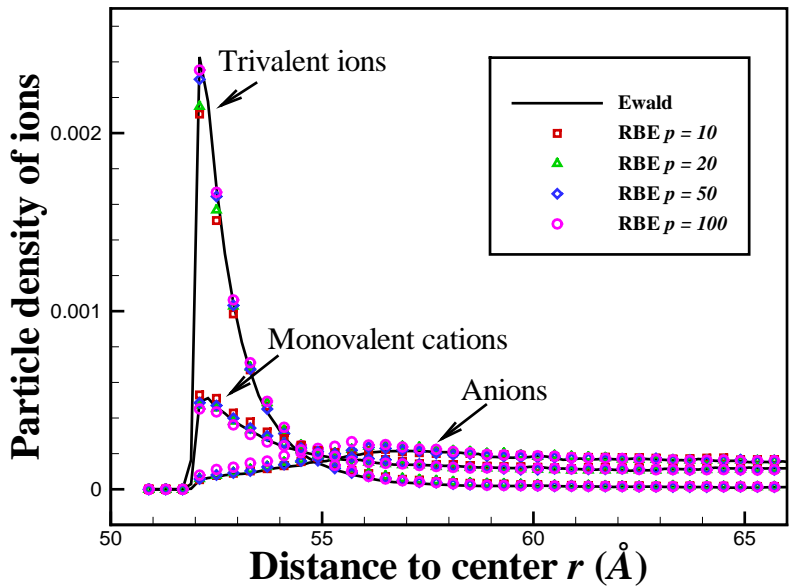}
	\caption{Contribution of different ion types to particle density $\rho$ when $c_{\mathrm{salt}}=196~\mathrm{mM}$.}
	\label{fig:chargeden196}
\end{figure}

The charge densities of different kinds of ions are shown in Figs. \ref{fig:chargeden0} and \ref{fig:chargeden196}, for $c=0~\mathrm{mM}$
and $c=196~\mathrm{mM}$, respectively. Clearly,  the RBE method can compute the densities correctly with acceptable accuracy. %However, the numerical heating due to random batch is still noticeable, especially around the peak of the density for $+3$ charged ions.
Again, larger batch size results in smaller errors. Anyhow, the simulation results seem to be acceptable here for all batch sizes.
For these small systems, the Andersen thermostat can already do a satisfactory job, and other temperature preserving techniques  can be considered for applications with large systems.

Overall, according to the two numerical examples above, we find that the RBE method is both accurate and efficient: it can correctly
capture the desired physical phenomena while saving the computational time significantly. We also expect the RBE easy to parallelize,
and to have good compatibility with a large number of cores. This will be tested in our subsequent work.
The extra noise introduced by the random batch can introduce some noticeable numerical heating. One may resolve this by increasing the batch size $p$ or decreasing the time step size to decrease the variance. Some more advanced techniques
include subtracting the effective temperature increase or use better thermostats like the Nos\'e-Hoover. Systematic studies of these will be in our subsequent work, too.

\section{Conclusions}\label{sec:conclusion}

In summary, we have developed a novel molecular dynamics method for particle systems with long-range Coulomb interaction using a RBE method
which needs only $\cO(N)$ operations in each step. The RBE method benefits from a random mini-batch idea
for the calculation of the force component in the Fourier series together with an importance sampling
for the Fourier modes. We have shown that the algorithm is accurate and efficient by calculating the correlation
functions between ions and the charge inversion near the colloidal surface by using two application examples
and demonstrated the promising properties for broader applications of the algorithm.
Besides the Coulomb systems, the RBE method can be extended to solve other many-body problems such as celestial bodies
and complex networks where the long-range interactions also play important roles.

It is remarked that our exploration of the RBE method so far is limited to a few aspects and there are much more work to
do in the future. If the system is partially periodic in some directions with Directlet or dielectric interface conditions
in other directions (e.g., the slab geometries), we believe the extension of our method is straightforward by introducing
techniques developed for such problems (see \cite{tyagi08,LiangYuan2020JCP,MaxianPelaez2021arxiv} and reference therein).
 In this work, we have compared the RBE with the classical Ewald summation, and this is actually
not fair for the demonstration of the CPU time. More reasonable comparison should be done through the PPPM which
is used in many implementations, and this comparison should be performed systematically in addition to
the demonstration on the scalability performance in parallel computing.
Also, the
simulations of this work are based on the primitive model of solvent. This model is simpler by treating water as
a continuum medium. This model is very good for the aim of numerical tests of electrostatic algorithms, but
the implementation for all-atom simulations shall generate much broader interest for practical uses.
All these issues shall be studied in our subsequent works.

\section*{Acknowledgement}
The work of S. Jin was supported by NSFC grant No. 12031013.
The work of L. Li was partially sponsored by NSFC 11901389, Shanghai Sailing Program 19YF1421300 and NSFC 11971314.
The work of Z. Xu and Y. Zhao was partially supported by  NSFC (grant Nos. 12071288 and 21773165) and the HPC center of Shanghai Jiao Tong University.
All the authors are supported by Shanghai Science and Technology Commission  (grant No. 20JC1414100).

\appendix

\section*{Appendix}

\section{The Debye--H\"uckel theory}\label{app:linearpb}

Consider an electrolyte solution with $N$ ions contained in the cubic box with PBCs, which are idealized as hard spheres of diameter $a$ and carrying charge $\pm q$. The numbers of anions and cations are both $N/2$ to meet electroneutrality condition. Let us fix one ion of charge $+q$ at the origin $\bm{r} = 0$ and consider the charge distribution around it.

 Inside the region $0<r<a$ there are no other ions, so the electrostatic potential satisfies the Poisson equation $\varepsilon\nabla^2 \phi=-q\delta(\bm{r})$ in this regime, where $\varepsilon$ is the permittivity of the solution. Outside this region, the charge of the $j$th species can be described by the Boltzmann distribution:
$\rho_j(\bm{r})=q_j\rho_{\infty,j}e^{-\beta q_j\phi}$ where $j=\pm$ and
$q_{\pm}=\pm q$, and $\rho_{\infty,+}=\rho_{\infty,-}=N/(2V)$.
Hence, when $r>a$:
\begin{align}
-\varepsilon\nabla^2 \phi=q\rho_{+}e^{-\beta q\phi}-q\rho_{-}e^{\beta q\phi}\approx \beta q^2\rho\phi,
\end{align}
which is the linearized Poisson-Boltzmann equation. By introducing the parameter $\kappa$ and Debye length $\lambda_D$ by
\begin{align}
\kappa\equiv\lambda_D^{-1}=\sqrt{\dfrac{q^2\rho}{\varepsilon k_B T}},
\end{align}
the solution of the Poisson equation can then be found to be:
\begin{equation}\label{eq:phiDH}
\phi(r)=
\begin{cases}
\dfrac{q}{4\pi \varepsilon r}-\dfrac{q\kappa}{4\pi\varepsilon(1+\kappa a)},& r<a,\\
\dfrac{qe^{\kappa a}e^{-\kappa r}}{4\pi\varepsilon r(1+\kappa a)},& r>a.
\end{cases}
\end{equation}

Hence, the net charge density for $r>a$ is given by
\begin{align}\label{eq:rhoDH}
\rho(r)=-\varepsilon\nabla^2\phi(\bm{r})=-\kappa^2\varepsilon\phi(r).
\end{align}
Obviously, $\rho(r)<0$ around the positive charge and
\[
\log(r|\rho(r)|)=-\kappa r+\log\left(\dfrac{\kappa^2 q\exp(\kappa a)}{4\pi(1+\kappa a)}\right)
\]
 is a linear function of $r$.
 The charge density around a negative charge is similarly discussed.

\section{Variance of the random force under the Debye--H\"uckel approximation}\label{app:forcevar}

We consider approximating the charge net density using the Debye--H\"uckel approximation to estimate,
\begin{equation}
\mathrm{Im}(e^{-i\bm{k}\cdot\bm{r}_i}\rho(\bm{k}))=\mathrm{Im}\left(\sum\limits_{j: j\neq i}q_j\exp(i\bm{k}\cdot(\bm{r_j}-\bm{r}_i))\right).
\end{equation}

We fix the ion $q_i$ at the center. For $r\ge a$, if we use the charge density $\rho$ given by \eqref{eq:rhoDH} and \eqref{eq:phiDH} to compute this quantity, we get
\begin{equation}\label{eq:DHapproxintegral}
\begin{aligned}
\sum\limits_{j: j\neq i}q_j\exp(i\bm{k}\cdot(\bm{r_j}-\bm{r}_i))&
\approx \int_{\R^3\setminus B(\b{r}_i, a)}\rho(\bm{r})e^{i\bm{k}\cdot\bm{r}}\,d\bm{r}\\
&=-q_i\dfrac{\kappa^2e^{\kappa a}}{2(1+\kappa a)}\int_{a}^{\infty}\int_{-1}^{1} r^2\dfrac{e^{-\kappa r}}{r}\cos(krz) dzdr\\
&=-q_i\dfrac{1}{1+\kappa a}\left[\dfrac{\kappa}{k}\sin(ka)+\cos(ka) \right]/\left(1+\dfrac{k^2}{\kappa^2} \right).
\end{aligned}
\end{equation}
This term is clearly real and the imaginary part is zero.
If one uses the DH theory to compute $\rho(\bm{k})$, one may get something bizzard in mathematics.
Using the same approximation leads to,
\[
e^{-i\bm{k}\cdot \bm{r}_i}\rho(\bm{k})
\approx q_i\left\{1-\dfrac{1}{1+\kappa a}\left[\dfrac{\kappa}{k}\sin(ka)+\cos(ka) \right]/\left(1+\dfrac{k^2}{\kappa^2} \right)\right\}.
\]
This means $\rho(\bm{k})\approx q_i e^{i\bm{k}\cdot \bm{r}_i}g(k)$ where $g(k)$ is independent of $i$. The left hand side does not depend on $i$ while the right hand side does. This clearly comes from treating all other particles except $i$ using the continuum approximation, and $i$ is not special in $\rho(\bm{k})$.  In spite of the bizzard result for computing $\rho(\bm{k})$, we believe that the approximation makes sense when one focuses on computing quantities associated with particle $i$, and $k\ll a^{-1}$.
When $k\ll a^{-1}$, the formula in \eqref{eq:DHapproxintegral} implies that $\mathrm{Im}(e^{-i\bm{k}\cdot \bm{r}_i}\rho(\bm{k}))=0$. This is understandable: in the equilibrium, provided that all other charges are distributed accurately by the continuum approximation, the net force is zero. In practice,  there is thermal fluctuation, and this cannot be zero, but it should be bounded by some number related to the temperature. Moreover, the magnitude of the integral on the right hand side is controlled by a bound uniform in $k$ (recall $|\sin(ax)/x|\leq a$) and we believe this result by continuum approximation can reflect the true magnitude of $\rho(\bm{k})$. Hence, when $k\ll a^{-1}$,  it is safe to bound $\mathrm{Im}(e^{-i\bm{k}\cdot \bm{r}_i}\rho(\bm{k}))$ by a constant.

Now,  if $\sqrt{\alpha}\lesssim a^{-1}$, we can then set $|\mathrm{Im}(e^{-i\bm{k}\cdot \bm{r}_i}\rho(\bm{k}))|\le C$ in computing \eqref{eq:varianceforce}. When $k\gtrsim \sqrt{\alpha}$, we do not assume the bounds on $|\rho(\bm{k})|^2$, as such terms will be dominated by $e^{-k^2/(4\alpha)}$. With $\sqrt{\alpha}\sim \rho_r^{1/3}$, one will have $S\approx
(\alpha L^2/\pi)^{3/2} \sim N$ by \eqref{eq:S}--\eqref{psf} so that
\begin{equation}
\begin{aligned}
\E |\bm{\chi}_i|^2 &=\frac{1}{p}\left(\sum_{\bm{k}\neq 0}\frac{(4\pi q_i)^2S}{V^2 k^2}e^{-k^2/(4\alpha)}|\mathrm{Im}(e^{-i\bm{k}\cdot\bm{r}_i}\rho(\bm{k}))|^2-|\bm{F}_{i,1}|^2 \right)\\
&\lesssim \dfrac{1}{p}\dfrac{(4\pi q_i)^2S}{V^2} \int_{2\pi/L}^{\infty} \left(\dfrac{L}{2\pi}\right)^3  \dfrac{4\pi k^2}{k^2}e^{-k^2/4\alpha}dk\\
&\approx \dfrac{8 q^2}{p}\dfrac{S}{V} \sqrt{\alpha\pi} \sim \frac{1}{p}\rho_r^{4/3}.
\end{aligned}
\end{equation}

%\bibliographystyle{siam} %{plain}
%\bibliography{ewald,groupbib,slab}

\end{document}